\documentclass[twocolumn,english,prl,amssymb,aps,superscriptaddress,showpacs,twocolumn,amsmath,showkeys,floatfix]{revtex4-2}
\usepackage[T1]{fontenc}
\usepackage{geometry}
\usepackage[active]{srcltx}
\usepackage{textcomp}
\usepackage{amsmath}
\usepackage{graphicx}
\usepackage{color}
\usepackage{hyperref}
\usepackage{esint}
\begin{document}


\title{Time-correlated Photons from a In$_{0.5}$Ga$_{0.5}$P Photonic Crystal Cavity on a Silicon Chip}

\author{Alexandre Chopin}
\affiliation{Thales Research and Technology, Campus Polytechnique, 1 avenue Augustin Fresnel, 91767 Palaiseau, France}
\affiliation{Centre de Nanosciences et de Nanotetchnologies, CNRS, Universit\'{e} Paris Saclay, Palaiseau, France}
\author{In\`es Ghorbel}
\affiliation{Thales Research and Technology, Campus Polytechnique, 1 avenue Augustin Fresnel, 91767 Palaiseau, France}
\author{Sylvain Combri\'{e}}
\affiliation{Thales Research and Technology, Campus Polytechnique, 1 avenue Augustin Fresnel, 91767 Palaiseau, France}
\author{Gabriel Marty}
\affiliation{Thales Research and Technology, Campus Polytechnique, 1 avenue Augustin Fresnel, 91767 Palaiseau, France}
\affiliation{Centre de Nanosciences et de Nanotetchnologies, CNRS, Universit\'{e} Paris Saclay, Palaiseau, France}
\author{Fabrice Raineri}
\email{fabrice.raineri@inphyni.cnrs.fr}
\affiliation{Centre de Nanosciences et de Nanotetchnologies, CNRS, Universit\'{e} Paris Saclay, Palaiseau, France}
\affiliation{Universit\'{e} Côte d'Azur, Institut de Physique de Nice, CNRS-UMR 7010, Sophia Antipolis, France}
\author{Alfredo De Rossi  }
\email{alfredo.derossi@thalesgroup.com}
\affiliation{Thales Research and Technology, Campus Polytechnique, 1 avenue Augustin Fresnel, 91767 Palaiseau, France}

\begin{abstract}
Time-correlated photon pairs are generated by triply-resonant Four-Wave-Mixing in a In$_{0.5}$Ga$_{0.5}$P Photonic Crystal cavitiy. Maximal efficiency is reached by actively compensating the residual spectral misalignment of the cavity modes. The generation rate reaches 5 MHz in cavities with Q-factor $\approx 4\times 10^4$, more than one order of magnitude larger than what is measured using ring resonators with similar Q factors fabricated on the same chip. The Photonic Crystal source is integrated on a Si photonic circuit, an important asset for applications in quantum technologies.
\end{abstract}

\maketitle

Parametric down-conversion, either through a second order or third order nonlinear optical process, underlies the emission of correlated photon pairs, entanglement \cite{kwiat1995} and squeezing. The miniaturization and the integration of sources based on these processes plays an essential role in quantum technologies \cite{grassani2015}. 
Maximizing the efficiency of the parametric interaction implies improved generation rate, entanglement, antibunching and the decrease of the pump power level \cite{steiner2021,ma2020}. Ultimate efficiency (i.e. single-photon nonlinearity) leads to deterministic quantum gates \cite{heuck2020}.
Considering resonantly enhanced pair generation through Spontaneous Four-Wave Mixing (SFWM), its rate R (or brilliance) depends \cite{azzini2012} on the Kerr nonlinear index $n_2$, on the Q-factor of the resonances, on the effective volume for the nonlinear interaction $V_\chi$ \cite{Marty2020} and on the pump power P delivered to the resonator, namely: 
\begin{equation}
R\propto n_2^2 \frac{Q^3}{V_\chi^2} P^2
\end{equation}
Therefore, the rate can be enhanced by design (increasing Q, decreasing the size of the resonator) or choosing a material with a larger $n_2$. Ring resonators made of AlGaAs, a group III-V semiconductor with sizable Kerr nonlinearity, have achieved a high (MHz) internal generation rate with 10 $\mu$W on-chip pump \cite{steiner2021}. On the other hand, due to scattering at sidewalls, the radius of high-Q rings made of semiconductors is hardly smaller than 10 $\mu$m. In contrast, high-Q photonic crystal (PhC) resonators with similar Q-factor confine the field in a volume which is orders of magnitude smaller \cite{asano2017}, thereby raising high expectations for a very strong nonlinear interaction. Yet, efficiency is reached only when the interacting "pump", "signal" and "idler" fields are simultaneously on resonance with the corresponding cavity modes. 
This condition, trivially satisfied in ring resonators, is extremely challenging in PhC. Triply-resonant SFWM was reported in a multi-mode resonator made of three coupled single-mode PhC cavities \cite{azzini2013}. The normalized generation rate (300 Hz $\mu$W$^{-2}$) is large, considering the moderate Q-factor ($\approx 5\times10^3$), owing to the small volume of the resonator. 
Correlated photon pairs have been reported in PhC waveguides \cite{xiong2011} and coupled cavities resonator waveguides \cite{Matsuda2013}, both with a pulsed pump and peak power level well above 100 mW.\\
A novel tuning technique, exploiting the inhomogenous thermo-optic effect, has enabled triply-resonant Four-Wave-Mixing (FWM) in multi-mode high-Q ($Q>10^5$) PhC, ultimately leading to the PhC Optical Parametric Oscillator \cite{Marty2020}. We show that this approach, here modified to operate with a fixed wavelength pump, enables time-correlated pairs are emitted at maximal efficiency (i.e. following the $Q^3$ scaling) within a PhC resonator, integrated with a Silicon Photonic circuit.
%
\begin{figure}[t!]
	\includegraphics[width=0.85\columnwidth]{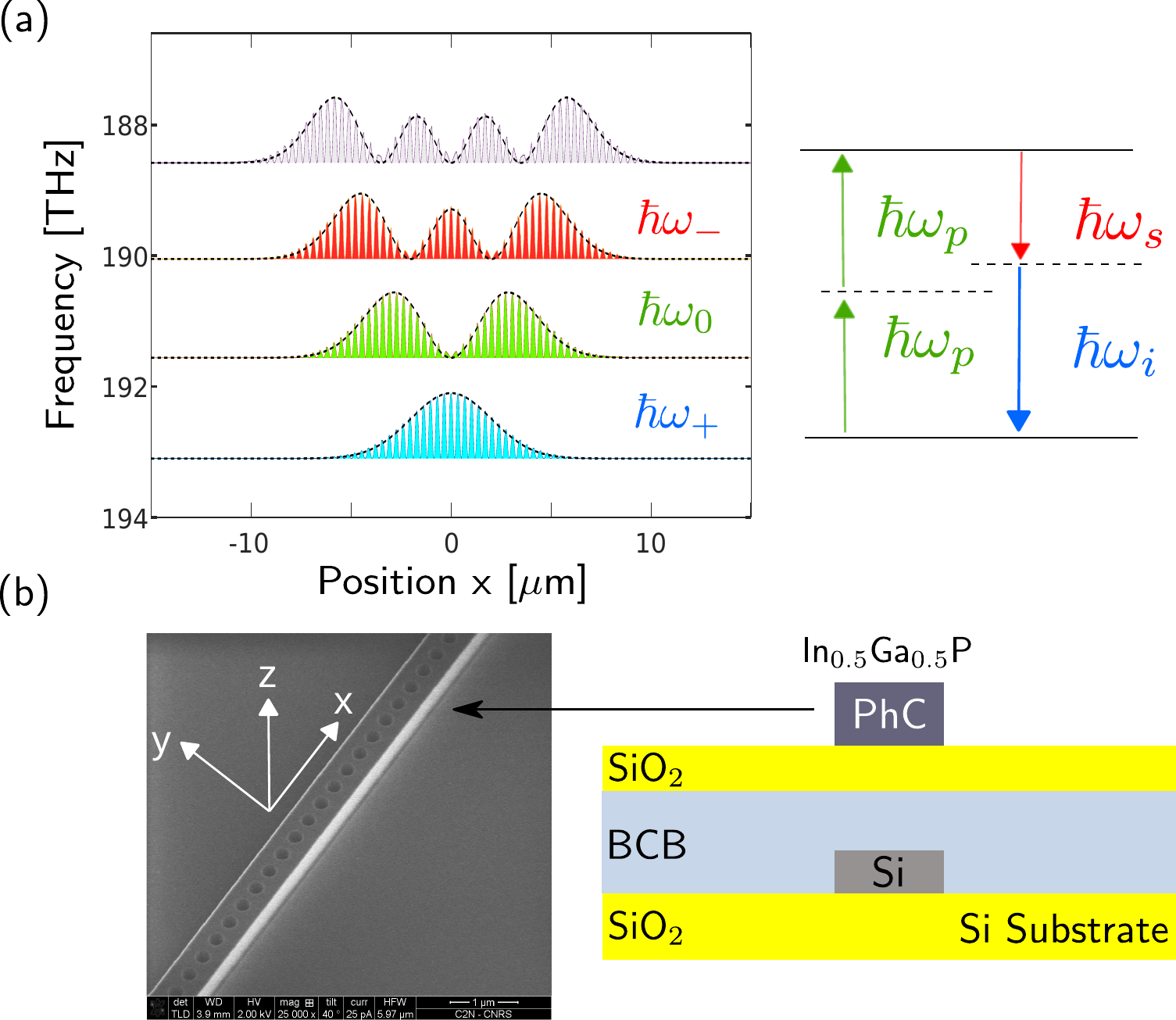}
	\caption{\label{fig:sample} Parametric PhC source. (a) Squared fields (calculated) of the first four modes (filled curves) of the PhC cavity and their envelopes (dashed lines) along the axis x, with $x=0$ the center of the beam; pump ($\omega_p$), signal ($\omega_s$) and idler ($\omega_i$) photons are resonant with modes $\omega_{0}$, $\omega_{-}$, $\omega_{+}$ in the SFWM process. (b) SEM image of the In$_{0.5}$Ga$_{0.5}$P PhC resonator on the Silica layer (left) and representation of the hybrid III-V PhC on Silicon layer stack (right)}
\end{figure}

The nanobeam PhC cavity \cite{Bazin2013} is made of In$_{0.5}$Ga$_{0.5}$P, a group III-V semiconductor alloy grown lattice-matched to GaAs. Its electronic bandgap (1.89 eV) is large enough to suppress Two-Photon-Absorption (TPA) of a pump in the Telecom spectral range \cite{combrie2009}. Thus, large electric fields can be established, enabling the observation of temporal solitons at chip scale \cite{colman2010}. Moreover, the residual linear absorption rate is about $1.5\times 10^{-8}$s$^{-1}$, which is extremely small, compared to other direct gap semiconductors \cite{ghorbel2019}.
The period of the holes is tapered to establish an effective harmonic potential for the field, involving nominally spectrally equispaced resonances at frequencies ($\omega_-$, $\omega_0$, $\omega_+$), Fig. \ref{fig:sample}(a). The PhC is coupled through evanescent field to an underlying Silicon photonic wire, Fig.\ref{fig:sample}(b). The full details on the fabrication and FWM process are given in Ref. \cite{Marty2019} and in the Supplemental Material. The cavities considered hereafter have quality factors ranging from $2\times10^4$ to $10^5$ and resonances between 192 THz and 194 THz (1545 nm - 1560 nm). The same chip contains ring resonators fabricated simultaneously with the PhC resonators, with a radius of 30 $\mu$m and comparable Q factors.\\ 
\begin{figure}[b!]
	\includegraphics[width=1\columnwidth]{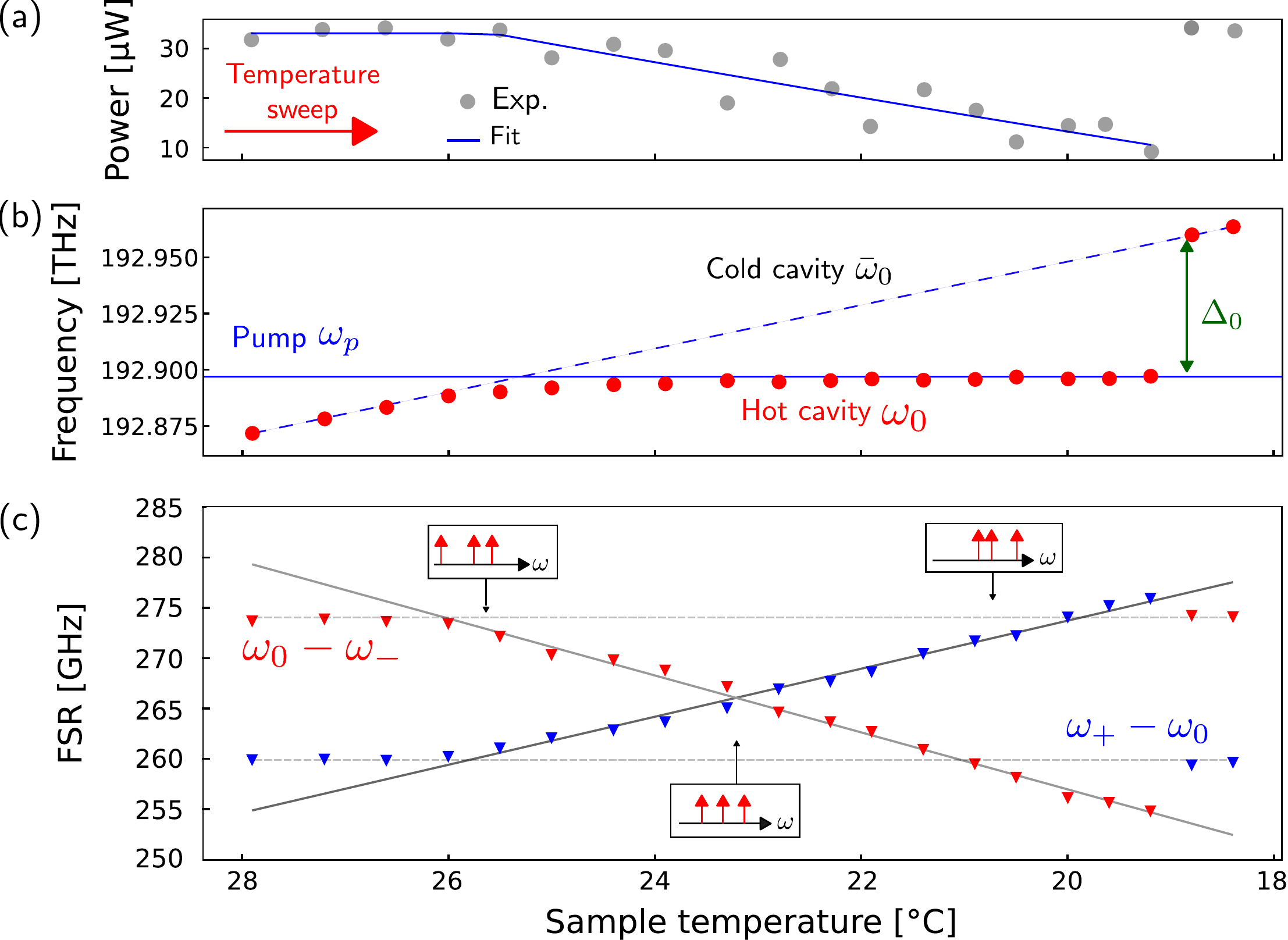}
	\caption{\label{fig:thermal_tuning} Achievement of equispaced resonances through thermal sweep and fixed pump wavelength. (a) Transmitted pump, after waveguide couplers (circles, line is a guide for the eyes) as the sample temperature is decreased; (b) corresponding resonance frequency $\omega_0$ (circles), solid and dashed lines represent the pump $\omega_p$ and the thermo-optic drift of $\overline{\omega}_0$ respectively; (c) corresponding measured FSR (markers), fit (solid lines) and value at uniform temperature (dashed lines); insets represent the aligned / not aligned configurations.}
\end{figure}
Compared to SFWM in a ring resonator, a specific feature here is the tuning process and the dependence of the emission rate on the injected pump power and the detuning. This stems from the markedly inhomogeneous and only partially overlapped distribution of the field of the interacting modes. Consequently \cite{Marty2020}, when the cavity heats up due to pump dissipation, the modes experience a differential thermo-refractive effect, which compensates for the unavoidable fabrication tolerances. 
%
\begin{figure*}[t!]
\includegraphics[width=0.95\textwidth]{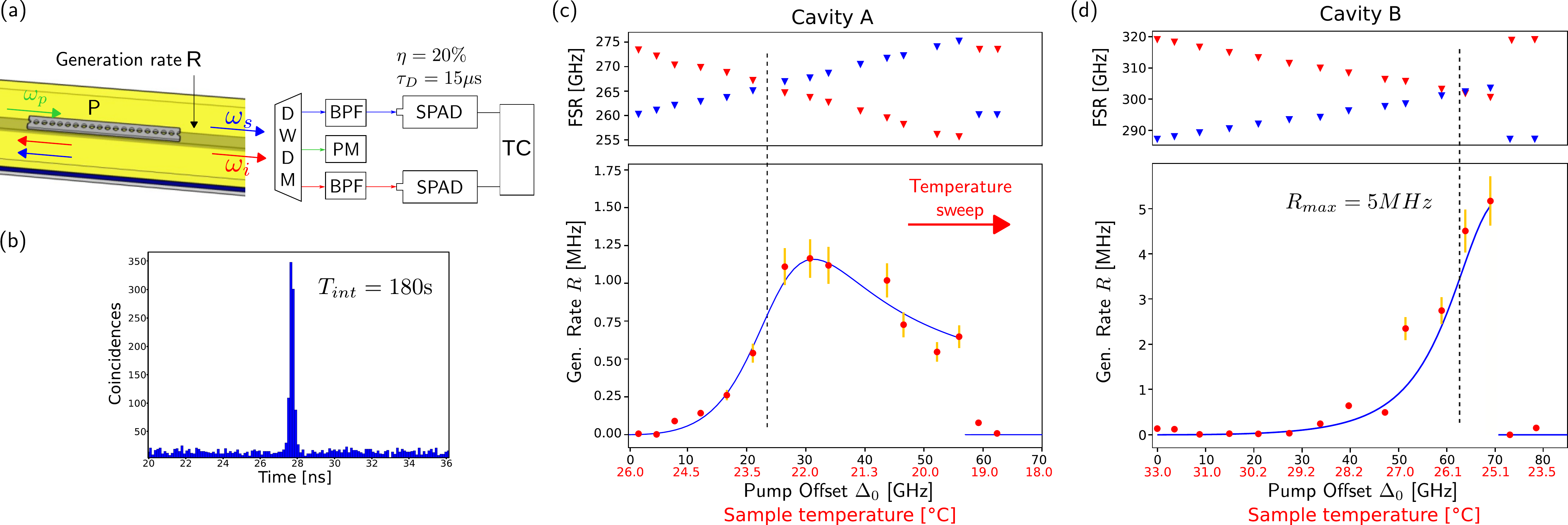}
\caption{\label{fig:pair-generation} Photon pairs generation in PhC. (a) Simplified time-correlation measurement set-up: Dense wavelength division multiplexing (DWDM), band pass filters (BPF), Power meter (PM), single-photon avalanche photodiode (SPAD), Time Correlator (TC). (b) Raw coincidence (at max. rate 5 MHz) histogram with integration time $T_{int}$, time bin 90 ps. (c,d)  On-chip pair generation rate $R$ (bottom) and corresponding FSR (top) as a function of the pump offset, controlled by the temperature; symbols and error bars are experimental points, solid line is theory with no fitting parameters; the vertical dashed line marks the equal FSR; (c,d) correspond to cavities ($A$,$B$).}
\end{figure*}
The pump laser is operated at a fixed frequency $\omega_p$, larger than the $\overline{\omega}_0$ resonance at sample temperature T=28°C, the overline symbol ( $\bar{}$ ) denotes hereafter the absence of internal dissipation, hence homogeneous temperature in the sample. The transmission decreases, as the temperature is decreased below T=26°C, until the bistable jump is reached at T=19°C, Fig. \ref{fig:thermal_tuning}(a). Optical Coherent Tomography (OCT) (see Supplemental Material, Ref. \cite{Combrie2017}) is used to track the corresponding thermal drift of the resonances, shown in Fig. \ref{fig:thermal_tuning}(b). As $\omega_0$ approaches the pump $\omega_p$, the energy stored in the mode, hence the internal dissipation, increase and $\omega_0$ is clamped to the pump (within 26°C - 19°C); the drift $\omega_0-\bar\omega_0$ is linearly related to the sample temperature, and can be approximated by the pump offset $\Delta_0 = \omega_p - \bar\omega_0 \approx \omega_0 - \bar\omega_0$. The differential thermo-optic effect is apparent in Fig. \ref{fig:thermal_tuning}(c), as the intervals $\Delta\omega_\pm=|\omega_{\pm} -\omega_0|$ change until they equalize at a given temperature (25°C here). 
Let us consider the mismatch $2\mathcal{D}=\Delta\omega_+ - \Delta\omega_-$. A well-defined temperature of the sample, at which the FSR are equalized and FWM is triply resonant, will exist only if the mismatch at uniform sample temperature is negative ($2\overline{\mathcal{D}}<0$ ). This is related to the sign of the differential thermo-optic effect. If this condition is satisfied, the pump power needs only to be large enough to keep the resonance locked until triple resonance is reached.\\ 
The fixed pump frequency is chosen such that signal and idler are both aligned with the channels of the DWDM (dense wavelength division multiplexing) filter  at the output of the sample, Fig. \ref{fig:pair-generation}(a). A bandpass (BP) filter is also used to remove the amplified spontaneous emission from the pump laser before entering the sample. Considering that the on-chip pump power P is roughly 250 $\mu$W and that most of it is reflected back on resonance (side-coupled cavity geometry), the 110 dB overall pump suppression at the detectors, ensured by cascading a BP filter on the signal and idler channels, is enough for our purpose. Two single-photon avalanche photodiodes (SPAD) (quantum efficiency $\eta_q$ = 20 \%, deadtime $\tau_D$ = 15 $\mu$s) are operated in gated mode (signal frequency f = 3 MHz, duty cycle $\sigma$ = 0.5) and connected to a time correlator (TC), as shown in Fig. \ref{fig:pair-generation}(a). The count of single photons on each channel ($N_1$,$N_2$) increases by orders of magnitude above the background simultaneously in both channels as the triply resonant FWM is established (see Supplemental Material).\\
A raw coincidences histogram, Fig. \ref{fig:pair-generation}(b), is collected as a function of the sample temperature with integration time $T_{int}$. The raw count of coincidences $C_{raw}$ (the sum of all coincidences over the main peak of width 2$T_j=300$ ps, $T_j$ the timing jitter of the detectors) is corrected with the estimate of the accidental counts $A$ (summing over the same width outside the peak) to generate the "true" coincidences $C_T=C_{raw}-A$. The Coincidence to Accidental Ratio is $CAR= C_T/A$, following Ref. \cite{guo2017}. To cope with the saturation of the SPADs as the count rate increases, further correction is applied and real coincidences are $C = C_T(1-N_1\tau_D)^{-1}(1-N_2\tau_D)^{-1}\sigma^{-1}$ \cite{hadfield2009}. Finally, the coincidence rate is then $R_{det}=C/T_{int}$ (see Supplemental Material)\\\
The on-chip pair generation rate $R$, that would be relevant for instance in a quantum chip, is deduced by taking into account the photon loss in the two channels $\alpha_{i}$ and $\alpha_{s}$ from the Silicon wire all the way to the detectors considering their quantum efficiency \cite{azzini2012}, hence $R = R_{det}/\alpha_{i}\alpha_{s}$. The attenuation ranges between 25 dB and 30 dB, depending on the sample. The main contributions are the tunable BP filters, the chip to fiber grating couplers and quantum efficiency of the SPADs. Moreover, in contrast with ring resonators, side-coupled PhC emit signal and idler photons with equal probability in both directions of the Silicon waveguide, giving an additional 3 dB loss.\\ 
%
Fig. \ref{fig:pair-generation}(c) shows the coincidence rate $R$ as the cavity is tuned towards its maximum efficiency by changing its baseline temperature, pump wavelength and power being fixed. A clear maximum, sharply emerging from noise is observed before a decrease due to a deviation from the triply resonant configuration. The sudden drop after the bistable jump clearly indicates that the system is then out of resonance. Yet, the maximum is reached after crossing the point of triple resonance, Fig. \ref{fig:pair-generation}(c,d). This is explained by a model is introduced hereafter. Starting from the connection between spontaneous and stimulated FWM \cite{liscidini2013}, the spontaneous emission rate is: 
\begin{equation}
R = \int {\eta_{\chi}d\nu}
\label{eq:pair_generation_rate_integral}
\end{equation}
the probability of stimulated conversion $\eta_{\chi}$ integrated over all the possible combinations of signal and idler frequencies satisfying the photon energy conservation. It can be expressed in the limit of undepleted pump and low parametric gain \cite{Marty2020}.
Combining this to eq. \ref{eq:pair_generation_rate_integral}, under the approximation that the intra cavity energy scales linearly with the pump offset (true if the absorption is linear), gives:
\begin{align}
R = \frac{\eta_{\chi}^{max}}{4}\frac{\Delta_{0}^2}{\Delta_{bist}^2}\frac{\Gamma_{-}\Gamma_{+}\left(\Gamma_{-}+\Gamma_{+}\right)}{\left(\Gamma_{-}+\Gamma_{+}\right)^2+16\overline{\mathcal{D}}^{2}\left(1-\frac{\Delta_{0}}{\Delta{opt}}\right)^2}
\label{eq:Pair_generation_rate}
\end{align}
where $\eta_{\chi}^{max}$ the maximum conversion probability at a given pump power, $\Gamma_{-,0,+}$ are the total cavity damping rate. $\Delta_{bist}$ ($\Delta_{opt}$) is the value of $\Delta_{0}$ for which the bistable jump (triply resonant configuration) is reached (see Supplemental Material).\\
The model (Eq. \ref{eq:Pair_generation_rate}) is superimposed to the measurement in Fig. \ref{fig:pair-generation}(c), with all these parameters being measured independently (see Supplemental Material) and none fitted. By inspecting the equation we note that $R$ depends on $\Delta_0$ via two terms: one is maximized when $\Delta_0=\Delta_{opt}$, i.e. triply resonant FWM, the other increases with $\Delta_0^2$. So, $R$ continues to increase although the triply resonance is loosely satisfied ($\Delta_0-\Delta_{opt}<\Gamma$). The agreement with the experimental points is remarkable. The model predicts that the optimal choice of the pump power is such that the bistable jump occurs just a little after the triply resonant FWM. This condition is shown in Fig. \ref{fig:pair-generation}(d), using cavity B, where a 5 MHz maximum on chip generation rate is estimated with roughly the same on-chip power as cavity A. Crucially, it demonstrates that the device can be tuned to its maximum efficiency which ultimately depends on the Q factor.\\
%
%
\begin{figure}[t!]
\includegraphics[width=0.83\columnwidth]{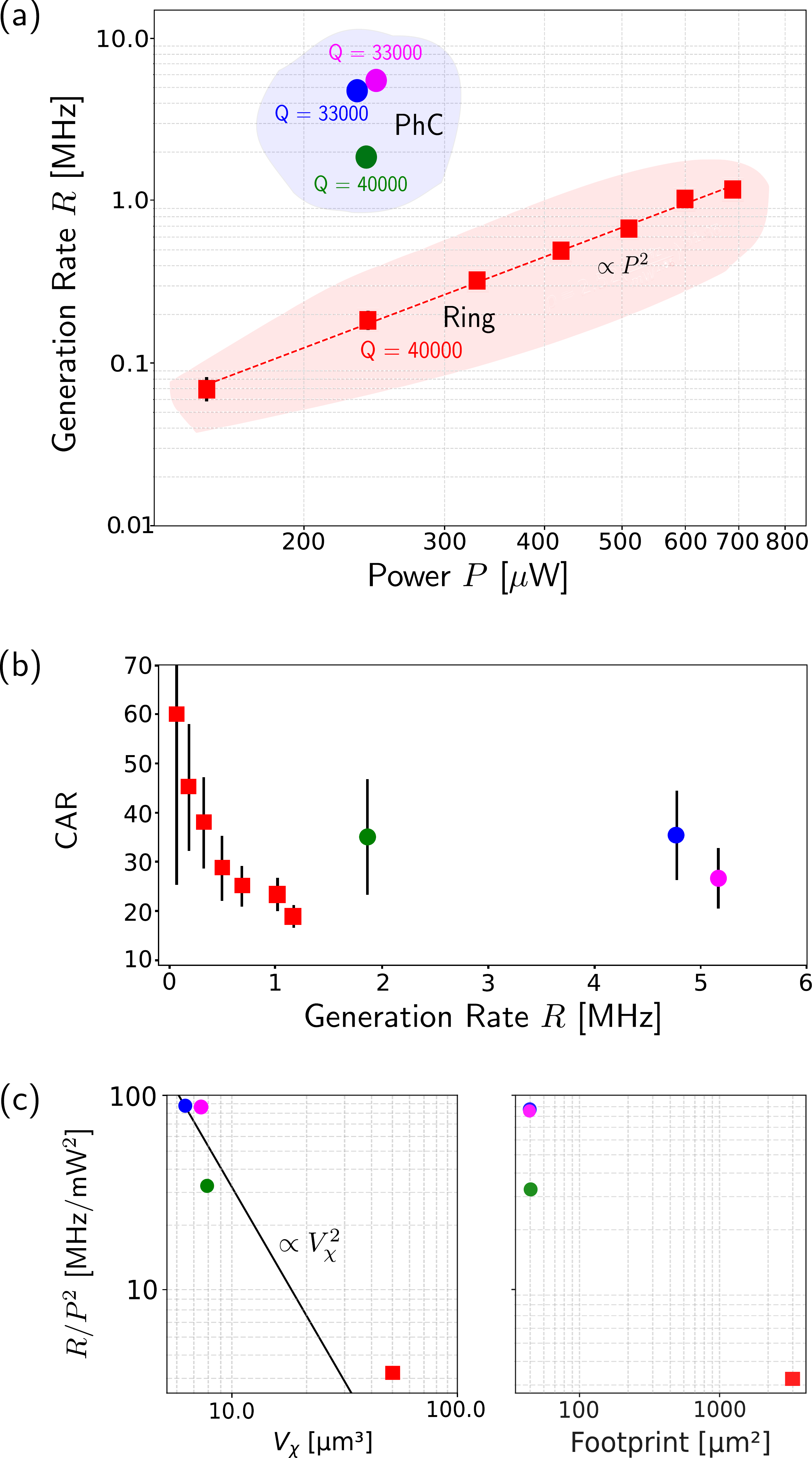}
\caption{\label{fig:PhCvsRing} SFWM in PhC (circles) and ring (squares) resonators. (a) On-chip pair generation rate vs. power (markers) and $P^2$ fit (dashed line); (b) corresponding CAR vs. generation rate (same color and symbol code); (c) Generation rate on chip normalized with pump power vs. the nonlinear interaction volume $V_{\chi}$ and the footprint of the device. Circles (squares) denote PhC (rings).}
\end{figure}
Let us compare the PhC with a ring resonator. The radius is $30 \mu$m, the same as in Ref.\cite{steiner2021}, and the  
Q factor is roughly the same as in the PhC (see Supplemental Material). As shown in Fig. \ref{fig:PhCvsRing}(a), the generation rate R scales with the square of the pump, as expected. Yet, it is apparent that for PhC cavity made of the same material, R is much larger by at least one order of magnitude, when normalized to the pump power. Fig. \ref{fig:PhCvsRing}(b) shows that in rings, the CAR decreases with the generation rate, indicating that accidental counts are mostly due to lost pairs, hence insertion loss \cite{guo2017}. The corresponding measurement for PhC reveals that the brillance is much larger at the same CAR. The generation rate is larger for PhC because of a much tighter confinement in PhC. This is apparent in Fig. \ref{fig:PhCvsRing}(c), as the maximum generation rate, normalized to the pump power, scales as the square of the inverse of $V_\chi$, which is an order of magnitude larger in the rings. Let us note that $V_\chi$ takes into account the spatial distribution and mutual overlap of the interacting modes, therefore it is distinct from the physical volume of the resonator; indeed, the difference in terms of footprint is even more pronounced.\\
In summary, we have reported time-correlated pair generation from a PhC resonator systematically operated at its maximum efficiency, owing to a thermal tuning mechanism. This only requires a fixed wavelength laser pump, removing the need for a tunable source, which is convenient for integration. The agreement with theory is within experimental error, with no need of fitting parameters. Simple scaling with volume is demonstrated by comparison with a ring resonator fabricated on the same chip. As the generation rate $R \propto Q^3$, we extrapolate a substantial (two orders of magnitude) improvement when using resonators with $Q=2\times10^5$ as for the recently demonstrated PhC OPO \cite{Marty2020}. This would essentially match the very recent achievement of ultra-bright source made of AlGaAs \cite{steiner2021}. The integration of the PhC with a photonic circuit leaves the possibility to include the essential functions of pump suppression and signal/idler separation, and in perspective, to operate several of such sources simultaneously.\\
This work was Funded by the French National Research Agency (ANR) under the contract COLOURS (ANR-21-CE24-0024).
The authors thank Gr\'{e}gory Moille for the design of the rings used here. Thierry Debuisschert and Kamel Bencheikh for discussions and clarifications. Laurent Labont\'{e} and S\'{e}bastien Tanzilli for having provided assistance in preliminary experiments in their facilities. See Supplemental Material at [URL will be inserted by publisher] for the derivation of the theory and for details about the experiments.

\bibliography{QuantumPhC_biblio}
\end{document}